# Formation of Nickel-Platinum Silicides on a Silicon Substrate: Structure, Phase Stability, and Diffusion from Ab initio Computations

M. Christensen[1], V. Eyert[1], C. Freeman[1], E. Wimmer[1], A. Jain[2], J. Blatchford[2], D. Riley[2] and J. Shaw[2]

[1]*Materials Design, Inc., Santa Fe, New Mexico, USA*

[2]*External Development and Manufacturing, Advanced CMOS, Texas Instruments Incorporated, 13121 TI Boulevard, MS 365, Dallas, Texas 75243, USA*

## Abstract

The formation of Ni(Pt)silicides on a Si(001) surface is investigated using an *ab initio* approach. After deposition of a Ni overlayer alloyed with Pt, the calculations reveal fast diffusion of Ni atoms into the Si lattice, which leads initially to the formation of $Ni_2Si$. At the same time Si atoms are found to diffuse into the metallic overlayer. The transformation of $Ni_2Si$ into NiSi is likely to proceed via a vacancy-assisted diffusion mechanism. Silicon atoms are the main diffusing species in this transformation, migrating from the Si substrate through the growing NiSi layer into the $Ni_2Si$. Pt atoms have a low solubility in $Ni_2Si$ and prefer Si-sites in the NiSi lattice, thereby stabilizing the NiSi phase. The diffusivity of Pt is lower than that of Ni. Furthermore, Pt atoms have a tendency to segregate to interfaces, thereby acting as diffusion barriers.

# I. Introduction

Formation of low Schottky Barrier Height (SBH) contacts for the source and drain regions of a MOSFET is critical to transistor performance and circuit speed. Although the strategy for such contacts is silicidation of a deposited metal layer, the choice of metal has undergone a change over recent nodes. This has been an imperative due to the inability of all metal silicides to exhibit low sheet resistance as line widths scale down. Recent nodes have required the progression to nickel silicide with the addition of platinum as an alloying constituent. The Ni-Pt-Si is a complex system in terms of the stability and properties of the various phases. Integrating such a system in to the CMOS fabrication flow depends on a good understanding of the temperature dependence of each phase and the process window available at each step in the flow for the formation of the desirable phase and inhibition of an undesirable phase. The role of the alloying constituent is to open up process windows and also to reduce the propensity for defect formation while maintaining low SBH. The contact formation process consists of two distinct thermal cycles. The first cycle is performed to allow compound formation in the regions where electrical contact is desired while inhibiting reaction in other regions so that metal may be easily removed from the latter regions but these requirements are in conflict and result in an uncertain process widow. When this is achieved it is at the expense of desirable electrical properties. Accordingly, after unreacted metal is stripped, the second thermal cycle must be performed in order to establish satisfactory electrical properties. Again, due to the fact that more than one phase with widely varying electrical properties may form, there is a narrow process widow for the desirable phase. The role of an alloying constituent is to open up the process window while reducing defectivity but this role has hitherto been poorly understood.

The ability to design optimal thermal cycles and take full advantage of advanced annealing techniques including laser annealing and alloying constituents depends on being able to model the kinetics of phase formation and evolution. This requires a calculation of activation energy values and diffusion parameters from fundamental considerations. With such information it is then possible to accelerate the time to establish an optimal process by predicting optimal temperature-time profiles at each step and using alloying constituents to maximum advantage.

Nickel silicides became an active research area in the 1970's thus providing a foundation for the silicide technology in the MOSFET fabrication process which was introduced in the 1980's [1][2]. Today, silicides remain an area of interest for technical development [3]. In particular, nickel silicide has taken on new importance for advanced technologies beyond the 65 nm technology node [4]. The push below the 28nm technology aggravates the importance of detailed materials control.

It is technologically advantageous to add noble metals such as Pd and Pt to the silicides. While the formation of pure nickel silicide is a well understood phenomenon, the presence of other elements complicates the process due to their differences in solubility and diffusivity in the various nickel silicide phases and the possibility of the formation of additional phases. For these reasons, it is desirable to establish a clearer atomistic understanding and a quantitative assessment of the structural and thermodynamic stability of the relevant silicide phases in the presence of alloying elements. This is the main objective of the present work and is accomplished by using accurate *ab initio* quantum mechanical calculations as described below. Furthermore, exploratory *ab initio* molecular dynamics simulations have been performed to gain insight into the dynamics of the silicidation process and the behavior of Pt.

For electronic device applications the monosilicide NiSi is the most important of the nickel silicides. Its resistivity of 14-20 mW-cm is comparable to that of $TiSi_2$ and $CoSi_2$, but NiSi can be formed at lower temperatures, namely 350-750 °C [5]. Also, less silicon is consumed to form this silicide (only 1.82 nm of Si is consumed per nm of metal) [5].

Although the nickel-silicon system has many phases, only three of those dominate the silicidation process, namely $Ni_2Si$, NiSi, and $NiSi_2$ [4]. Empirically it has been found that silicidation can be accomplished by the following steps. After deposition of Ni on silicon, $Ni_2Si$ is formed in the temperature range of 250–300 °C. This nickel rich phase rapidly transforms around 300 °C into the low resistivity NiSi film. However, if the temperature is too high, $NiSi_2$ will appear resulting in an undesirable higher electrical resistivity [4].

For (near) noble-metals (Ni, Pd, Pt) silicides, metal atoms are the main diffusing species in the growth of the first phase, whereas for refractory metals silicon is the diffusing species [6]. From marker experiments, it has been observed that for metal-rich silicides such as $M_2Si$, the dominant diffusing species are mostly metal atoms. On the other hand, in the formation of monosilicide and disilicide, silicon atoms are reported to be generally the dominant diffusing species [7]. In typical processes with 10 nm thick Ni films, the metal film is completely consumed resulting in a film thickness determined by the deposited metal thickness. The process relies on Ni diffusion into adjacent silicon, but Ni deposited on top of $SiO_2$ isolation areas and $Si_3N_4$ spacers is not consumed but migrates toward poly lines and source/drain silicon regions [4]. For Ni and Pt, the second phase, the monosilicide, does not form until all the metal is consumed in forming the first phase. Consequently, structures can be consistently formed with either the first or second phase next to the silicon [6].

Chang and Erskine used photoemission and work function measurements to investigate the formation and structure of Ni-Si(100) interfaces at 300 K [8]. They found that for Ni coverages $\theta \leq 0.5$ Å ($4.6 \times 10^{14}/cm^2$) a chemisorbed phase of Ni surface atoms forms. This chemisorbed phase persists up to coverages of $\theta \approx 2$ Å; but also in this coverage range, a diffusion layer appears in the Si lattice. At $\theta \approx 1.5$ Å the surface composition closely resembles NiSi. Addition of more Ni atoms ($\theta \geq 1.5$ Å) initiates nucleation of $Ni_2Si$. The growth of this phase continues up to $\theta \approx 1.5$ Å where silicide formation stops. Additional Ni atoms deposit as a pure Ni overlayer. These observations yield a model for Ni-Si interfaces formed at room temperature which consists of a shallow diffusion layer of Ni atoms in the Si lattice, a very thin (1.5 Å thick) interface having the chemical characteristics of NiSi, a $Ni_2Si$ phase (15 Å thick) of relatively uniform stoichiometry followed by a Ni layer. While providing valuable insight, photoemission experiments are limited to the top-most surface layers and do not yield information of thicker silicide films.

Erokhin *et al.* [9] observed formation of continuous 50 nm thick polycrystalline silicide layers by thermally induced reaction of a Ni film deposited on a preamorphized Si substrate at a temperature as low as 360 °C. The silicide layer was completely confined within the 65 nm thick amorphous layer formed by Si+ ion implantation. After first annealing at 360 °C for 8 hours, all the Ni atoms reacted with amorphous silicon, forming mostly a $NiSi_2$ phase, where resistivity was found to be 44 $\mu\Omega$ cm. Some non-reacted amorphous silicon was left within the original amorphous to crystalline interface due to lack of Ni to consume all the amorphous silicon. An additional annealing step at 400 °C for 1 hour leads to complete epitaxial recrystallization of the residual amorphous regions due to lateral Ni disilicide growth. Ni atoms necessary to complete the reaction are provided due to partial decomposition of the already formed $NiSi_2$ layer. But even in this case, no silicide was found beyond the original a-Si layer.

The addition of Pt is known to increase the thermal stability of NiSi [10]. Pt has a distinct influence on the texture and stability of NiSi on Si [11]. The expansion of the NiSi lattice caused by the incorporation of Pt is thought to be the origin of the texture evolution. Cojocaru-Mirédin *et al.* studied the redistribution of Pt after heat treatment at 290 °C for 1 h by large-angle atom probe tomography assisted by femtosecond laser pulses [12]. Two silicides $Ni_2Si$ and NiSi were found together with the solid solution of Ni and Pt. The redistribution of Pt at the $Ni_{1-x}Pt_x/Ni_2Si$ interface was attributed to the snowplow effect (rejection of an alloying element during phase growth where the alloying element is pushed away by a moving interface) and segregation of Pt at the $Ni_2Si$/NiSi interface was observed.

Demeulemeester *et al.* studied the redistribution of Pt during nickel silicidation using real-time Rutherford backscattering spectrometry [13]. The Pt redistribution was found to be inhomogeneous, which was attributed to the low solubility of Pt in $Ni_2Si$ compared to NiSi and the limited mobility of Pt in NiSi. Pt acts as diffusion barrier and resides in $Ni_2Si$ grain boundaries, slowing $Ni_2Si$ and Ni growth kinetics. Increasing Pt concentration raises the temperature of complete silicide formation.

Hoummada *et al.* also studied the redistribution of Pt during the silicide formation [14]. During the normal growth of $Ni_2Si$, Pt is retained in the Ni-Pt layer. This is mainly due to the low solubility of Pt in $Ni_2Si$ and possibly due to the low diffusion of Pt in $Ni_2Si$. This solubility has been measured to be about 8% at 1073 K and is estimated to be about 1% at 573 K by thermodynamics calculations. After the complete formation of NiSi at low temperatures, Pt is located in the middle of the NiSi layer. Simultaneous growth of $Ni_2Si$ and NiSi during the *in situ* annealing was observed. It was attributed to the slowdown of the formation kinetics of $Ni_2Si$ by the accumulation of Pt at the interfaces Ni-Pt /Ni$_2$Si and Ni$_2$Si/ NiSi acting as diffusion barriers for Ni.

While these experimental data provide an overall phenomenological picture of the silicidation process, a detailed atomistic description and understanding is still incomplete. For example, the chemical potential of Pt atoms in the various nickel silicide phases and the thermodynamic driving forces for segregation to interfaces, hence, the distribution of Pt atoms and their kinetics, are not fully established. An interesting experimental technique to address the spatial distribution of an element on the atomic level in a material is atom probe tomography. Local electrode (3D) atom probe measurements have indeed been carried out on Pt together with other dopants (Pd [15], As and B [16]) in Ni$_{1-x}$Pt$_x$Si silicide films. It was found that Pt atoms segregates to the silicide/Si interface.

Atomic simulations can add valuable insight which may be hard to obtain from experiments, in particular the structure, energies, and electronic properties of buried interfaces. In fact, *ab initio* simulations have been used to study structure and energy of Si/NiSi$_2$ interfaces [17,18]. Due to the importance of nickel silicide as a contact material in CMOS devices, most of the computational studies address the electronic properties of the interface between silicon and nickel silicide, such as the Schottky barrier height in e.g. Refs. [19,20], and the influence of dopants on such properties [21,22]. A range of properties (structural, electronic, vibrational, and elastic) of pure bulk NiSi were obtained from a*b initio* simulations by Connétable and Thomas [23]. Studies have also been performed on Pt silicides and interfaces. Structural, electronic and surface properties of PtSi and Pt$_2$Si silicides have been studied [24] as well as the effect of Ti on the Si/PtSi Schottky barrier height [25].

Both the Ni-Si and the Pt-Si phase diagrams are rich in phases exhibiting a number of different crystallographic structures [26]. Ni forms silicides including $Ni_3Si$, $Ni_{31}Si_{12}$, $Ni_2Si$, $Ni_3Si_2$, NiSi, and $NiSi_2$. Pt silicides include phases with corresponding stoichiometry as in $Pt_3Si$, $Pt_2Si$, and PtSi. However, Pt silicides do not generally have the same crystal structure as Ni silicides even though the stoichiometry is the same. Some other differences between the Ni-Si and Pt-Si systems are also visible in the phase diagrams. Si has significant solubility in Ni, but not in Pt. Also, Pt has a higher melting point than Ni.

It is the purpose of the present computational work to obtain a clearer picture and a more detailed quantitative description of the key structural and thermodynamic properties of the Ni-Pt-Si system in the context of silicide formation on Si substrates. Additionally, *ab initio* molecular dynamics simulations have been carried out to shed light on the diffusion mechanism and the role of Pt in the silicidation process.

## II. Computational Approach

The silicidation process is approached from two different perspectives, namely (i) by detailed, quantitative calculations of the structural and thermodynamic properties of the relevant phases, *i.e.* metallic Ni(Pt), $Ni_2Si$, NiSi, $NiSi_2$, and Si; (ii) by a direct simulation of the silicidation process using *ab initio* molecular dynamics.

The starting point of the *ab initio* molecular dynamics simulations is a Ni(Pt) overlayer on a Si(001) surface. The Si phase is n-doped with As and P. The Ni phase contains about 10 % Pt. Therefore, a $Ni_{0.9}Pt_{0.1}$ (111) metal overlayer is deposited on an n-doped Si(001) surface. The atomic model contains three metal layers and 16 Si layers forming an interface. There are a total of 191 atoms in the supercell which is periodic in two dimensions. One silicon atom is substituted by P and two silicon atoms are substituted by As. Two different structures are considered, differing in the positions of the As and P dopant atoms. In model 2, the positions of P and the As atom closest to the interface are swapped. The atomic structure before and after annealing are shown in Fig. 1. The structures in Figs. 1e and 1f are used to investigate the propensity for Pt to segregate to the Si/silicide interface.

The evolution of the silicidation process is studied using *ab initio* molecular dynamics. The system is annealed at 2000 K (this temperature was selected to provide substantial configurational sampling) until the energy is becoming reasonably constant, *i.e.* the system has reached a local equilibrium at this temperature. The Si layers furthest away from the interface are kept frozen during the anneal. The system is subsequently cooled down to low temperature (T = 0 K), where the atoms are relaxed to their equilibrium positions. It should be noted that the practical limitations of *ab initio* molecular dynamics require a cooling rate which is by far faster than experimental conditions. Thus, only a relatively small part of the configuration space is explored. This limitation should be kept in mind in the interpretation of the present calculations. In contrast, the thermodynamic calculations including the chemical potential of Pt in the various Ni phases do not invoke kinetic effects.

All total energies and forces on the atoms are computed with *ab initio* density functional theory (DFT) [27,28]. Electron-electron exchange and correlation effects are described by the generalized gradient approximation (GGA) with the PBEsol functional form [29], which is a revised version of the functional proposed by Perdew-Burke-Ernzerhof (PBE) [30]. The density-functional or Kohn-Sham equations are solved with projector-augmented-wave (PAW) potentials [31] as implemented in the Vienna *ab initio* Simulation Package (VASP) version 5.2 [32-35] within the MedeA® computational environment [36]. The Brillouin zone is sampled with a k-point mesh with a density of at least 0.5 Å$^{-1}$, and the plane-wave cutoff energy for expansion of the wave functions is set to 269.5 eV.

## III. Results and Discussion

### A. Structural Features of Nickel Silicides

The structural properties of nickel silicides are well described by the present *ab initio* calculations. Computed lattice constants compared with experimental values are given in Table I, for $Ni_3Si$, $Ni_{31}Si_{12}$, $Ni_2Si$, $Ni_3Si_2$, NiSi, and $NiSi_2$ together with those of Si and Ni. The deviation between computed and experimental structural parameters is less than 2 % for all phases which is the standard present level of theory. The results show that the Ni-Ni bond distances are too short while the Si-Si bond is slightly too long. Hence, as a side-effect, the deviation of the lattice parameters from experiments diminishes with increasing Si concentration.

The atomic structure of nickel silicides and the structural relation between different silicide phases give valuable information about silicidation and the transition between silicides. NiSi has an orthorhombic crystal structure in the space group *Pnma* with computed lattice parameters of *a* = 5.08 Å, *b* = 3.36 Å, and *c* = 5.53 Å. The Si sublattice of NiSi can be viewed as a distorted lattice of the Si bulk crystal structure. The Ni atoms occupy distorted octahedral interstitial sites in the [110] channels of the silicon sublattice. The number density of Si atoms in NiSi is 41.1 atoms per nm$^3$, which is about 20% less than in pure bulk Si, as shown in Table II. In other words, NiSi can be viewed as an expanded and distorted Si lattice with Ni atoms filling interstitial sites.

A similar argument holds for $Ni_2Si$. This phase crystallizes also in an orthorhombic lattice with the space group *Pnma*. The computed lattice constants are $a$ = 4.91 Å, $b$ = 3.71 Å, and $c$ = 7.01 Å. The Si sublattice in $Ni_2Si$ has still the same topology as pure Si. However, the (001) planes are sheared. Ni atoms occupy two different sites in $Ni_2Si$. One type of the Ni sites is in the structure framework, similar to the sites occupied by Si atoms. The other type of sites is inside the channels of the Si lattice in the same way as in NiSi. The two types of Ni atoms are computed to be equally stable in the lattice.

Hence, NiSi can be formed by inserting Ni atoms into channels of a Si lattice accompanied by an expansion of the Si sub-lattice. However, the formation of $Ni_2Si$ requires atomic rearrangements which result in Si atoms being replaced by Ni atoms.

The number densities of Si, Ni, and nickel silicide phases are summarized in Table II. The highest Si density is found in $NiSi_2$ and the lowest in $Ni_3Si$, as expected. The volume of the Si sublattice is more than doubled from $NiSi_2$ to $Ni_3Si$. However, the total number density is increasing with higher Ni content. The volume per atom in $Ni_3Si$ is half of that in pure Si. This trend is consistent with the pure systems: silicon has about half the number of atoms per unit volume than nickel. Furthermore, the number density of nickel silicides is more like bulk Ni than bulk Si. The number density of Si in $NiSi_2$ is almost the same as in pure Si. This corroborates the view that the formation of $NiSi_2$ can be considered as filling of interstitials in the Si lattice by Ni atoms accompanied by a transition from covalent semiconductor to dense metallic silicide.

### B. Formation of Initial Nickel-Silicides

As stated earlier, the most important nickel silicide phases for the silicidation process are $Ni_2Si$, NiSi, and $NiSi_2$. During silicidation, these phases form at different temperatures with $Ni_2Si$ prevailing at 300 °C, NiSi at 400 °C, and $NiSi_2$ at 800 °C [37]. The stability of $Ni_2Si$ is reflected in its relatively high melting point of 1305 °C, which is not far below that of pure Ni (1455 °C) and pure Si (1414 °C). In comparison, the melting point of NiSi is 990°C while $NiSi_2$ decomposes at 966 °C into Si and NiSi.

Thermodynamic data, such as formation energies and chemical potentials, provide information on the driving forces for atomic rearrangements and phase formation during the silicidation process. Due to their high computational accuracy, *ab initio* calculations are well suited to obtain quantitative information of the energetics.

The computed reaction energies leading to the formation of silicides are given in Table III. The calculations show that the energy of formation of the Ni-rich phase $Ni_3Si$ is strongly exothermic,

$$\text{Si} + 3\,\text{Ni} \rightarrow \text{Ni}_3\text{Si} \quad \Delta E = -206.3 \text{ kJ/mol} \tag{1}$$

thus being a favorable reaction in a Ni-rich region. $Ni_3Si$ can continue to react exothermally with Si to form $Ni_2Si$

$$\text{Si} + 2\,\text{Ni}_3\text{Si} \rightarrow 3\,\text{Ni}_2\text{Si} \quad \Delta E = -138.2 \text{ kJ/mol} \tag{2}$$

albeit with a smaller thermodynamic driving force. It becomes increasingly less favorable for Si in a Ni rich environment to form silicides with increasing Si content, *i.e.* NiSi and $NiSi_2$, as shown in Table III. Hence, there is a very strong thermodynamic driving force for silicon to move into Ni and to form $Ni_2Si$ and also NiSi, but not so much $NiSi_2$.

There is also a thermodynamic driving force of Ni to move into Si and to form $NiSi_2$, which can favorable react with more Ni to form NiSi and $Ni_2Si$, but not so much $Ni_3Si$. This means that for a Si surface with a nickel metal overlayer, Ni and Si interdiffusion is thermodynamically favorable leading preferentially to NiSi and $Ni_2Si$. This is consistent with experiment.

Ni atoms become less stable in the silicides when the nickel content increases. This is shown in Fig. 2, which displays the energy gain of inserting Ni or Pt atoms into a nickel vacancy in nickel silicides. This is the negative chemical potential and hence a measure of the stability of these elements in the silicides. The chemical potential of Ni is lowest in $NiSi_2$. That is, Ni is most stable in $NiSi_2$. The chemical potential of Ni increases with increasing Ni content in silicide, except for $Ni_3Si$. Generally, Ni is most stable when highly coordinated with Si. Pt follows the same trend as Ni. In almost all cases it is energetically most favorable to fill a vacancy by a Ni atom rather than by a Pt atom. The exception is NiSi. This is the silicide in which Pt has the highest stability. In other words, Pt stabilizes the monosilicide relative to the other silicides $Ni_2Si$ and $NiSi_2$.

Before considering the role of Pt and the distribution of P and As dopants in the silicides, we list in Table IV the computed energies of moving a Ni, Pt, P, and As atom from the elemental state into Si. The insertion of Ni, Pt, P, and As atoms into a silicon lattice is endothermic except for the substitution of Si atoms by P atoms. Remarkably, the substitution of a Si atom by a Pt atom is only slightly endothermic (5 kJ/mol) whereas the substitution of Si by Ni is energetically quite unfavorable (118 kJ/mol). On the other hand, interstitial Ni is much more stable relative to interstitial Pt. Including configurational entropy, Pt, P, and As can be thermodynamically stable at substitutional sites in crystalline silicon whereas Ni atoms can be stable at interstitial sites.

The energies required to remove Pt, As, and P from their respective elemental bulk phase and substituting a Ni or Si atom in nickel silicides are given in Table V. Pt prefers to substitute Ni in the silicides, in particular in the monosilicide NiSi. An energy of 40 kJ/mol is gained when Pt moves from $Ni_2Si$ to NiSi, swapping with Ni. Hence, the computations clearly show that Pt stabilizes NiSi. Pt substituting Si in the silicides is associated with a high energy cost. The exception is in NiSi$_2$, for which substitution of Si by Pt is highly favorable releasing 82 kJ/mol. A closer look at the specific atomic arrangements around the Pt atom in NiSi and in NiSi$_2$ shows that they are similar (Fig.3). The local atomic arrangement around Pt in NiSi$_2$ being somewhat more symmetric than in NiSi. Thus, there is a driving force for Pt to move from NiSi to NiSi$_2$ and thereby locally transform NiSi$_2$ to (Ni,Pt)Si.

The energy values in Table V show that P and As dopants are thermodynamically relatively more stable inside the silicon phase as compared to within the silicides. Hence, during the silicidation process, As and P prefer to stay in the Si phase and avoid becoming embedded in the silicide phase when the silicide grows. However, both P and As atoms are very stable at substitutional sites in metallic nickel. This may be interpreted as indicating that excess metal and high processing temperatures may deplete doped silicon of P and As. The stability of arsenic in metallic Ni is consistent with the formation and stability of NiAs (NiAs exists in nature as the mineral nickeline). Phosphorus and arsenic atoms incorporated into silicides would prefer to substitute Si atoms. Phosphorus has a higher propensity to be in the silicide than arsenic, especially in NiSi, and some P can be embedded in the silicide. Energy differences of As and P in the silicides and in pure Si are explicitly given in Table VI. Notably, the energy cost of moving P from Si into NiSi is only 11 kJ/mol. Hence, at elevated temperatures entropic effects could cause P to migrate into NiSi.

While computed chemical potentials and formation energies as given above provide valuable information about the thermodynamic driving forces during silicidation, the kinetic aspects of the silicidation process are investigated by annealing of Ni(Pt) deposited on a silicon substrate (Fig.1).

The structures after atomic relaxation by energy minimization are shown in Fig. 1a and 1b. These models represent (Ni,Pt) on Si before annealing. The interface has a Ni-Pt overlayer which approaches the Si surface closely. The metal overlayer undergoes a relaxation resulting in a warping of the Ni-Pt (111) film. The top three Si layers closest to the metal overlayer are strongly perturbed from their surface positions, and the Si surface dimers disappear. There is some intermixing of Ni and Si at the interface but it remains slight initially. The dopant atoms As and P remain at their Si sites. The energy difference between the models is small, model 1 being 5 kJ/mol lower in energy than model 2.

The anneal leads to substantial atomic rearrangements at the metal-semiconductor interface. Figs. 1c and 1d show the resulting atomic structures of the two models. In the annealed structures, there has been an interdiffusion of Si into Ni and of Ni into Si, leading to the formation of several layers of silicide. The silicide is semi crystalline. In model 1, phosphorus and one arsenic atom have been pushed down with the advancing Ni front, and remain in the Si phase. This snowplowing of the dopants by the advancing silicide front is consistent with the thermodynamic data in Tables V and VI. Snowplowing of As by the silicide formation has also been found experimentally [16]. The As atom closest to the interface has diffused through the silicide layers and becomes a surface atom on the silicide surface. However, this behavior is dependent on the initial structure. In model 2, phosphorus is initially located close to the interface but is embedded in the silicide after the anneal. This is consistent with the small energy preference of phosphorus in Si as compared to in the silicide. Also the As atoms are more in the mixed Si-Ni phase than in the pure Si phase. Thus, in simulations with rapid quenching, P and As atoms can be trapped inside the silicide layers.

Concentration profiles of Ni, Pt, Si, As, and P before and after anneal are shown in Figs. 4 and 5 for models 1 and 2, respectively. In both cases, a silicide layer about 10 Å thick is formed. The top three layers from the silicide surface have the composition of $Ni_2Si$. The next few layers have the composition of NiSi and $NiSi_2$. The $Ni_2Si$ layers are oriented with [100] parallel to the [001] direction of the Si substrate. The (100) surface is also the most stable $Ni_2Si$ surface (*cf.* Table VII). The fourth silicide layer from the silicide surface (the layer closest to the Si substrate) has the composition of NiSi. Comparing the atomic structure with that of NiSi shows that it is similar to NiSi viewed in the (010) direction. In fact, the (010) surface of NiSi is the most stable free NiSi surface (*cf.* Table VIII).

The boundary between silicide and the silicon substrate remains relatively sharp (a few atomic layers) due to the fact that the chemical potential of Ni in bulk Si is 45 kJ/mol (energy required to move Ni atoms from bulk Ni into interstitial sites in silicon) whereas the formation energy of NiSi from Si and bulk Ni is -115 kJ/mol. This means that there is a strong thermodynamic driving force for Ni atoms to form a NiSi phase rather than being dissolved in bulk silicon. This quantifies the observation of a very low solid solubility of Ni in Si as seen in the phase diagram.

Because of the high mobility of Ni in bulk Si and the attractive chemical potential of Ni inside NiSi the concentration of Ni atoms ahead of the silicide front is determined essentially by thermodynamics, even at low temperatures in the processing. With a difference in the chemical potential of 160 kJ/mol, even at 1000 °C the equilibrium concentration of Ni in Si would only be about $7 \times 10^{-7}$ %. Upon cooling of such a sample Ni atoms would continue to diffuse from the interior of Si towards the silicide phase, and thereby reducing the amount of dissolved Ni in Si even further.

The diffusion of Ni into Si occurs faster and is more pronounced than diffusion of Si into the metal phase in the initial stage forming $Ni_2Si$. Hence, Ni is the dominant diffusing species during the formation of $Ni_2Si$. In comparison, Co forms silicides in the same sequence as Ni, *i.e.* $Co_2Si$ → CoSi → $CoSi_2$. Co is reported to be the dominant diffusing species during $Co_2Si$ formation while Si is the dominant diffusing species during the CoSi formation [38]. Thus, the simulations indicate that the metal is the dominant diffusing species in the initial stage forming $Ni_2Si$, just as in the formation of $Co_2Si$. This is not surprising as the $Ni_2Si$ and $Co_2Si$ structures are isomorphic with symmetry *Pnma* with almost the same lattice parameters. On the other hand, while the monosilicide NiSi has a *Pnma* structure, CoSi has a cubic *P2_13* structure. In the transformation of $Ni_2Si$ to NiSi, Si is the main diffusing species as discussed below.

## C. Transformation of $Ni_2Si$ into NiSi

In the transformation process of $Ni_2Si$ to the desired NiSi, a number of mechanisms need to be considered. Overall, the progression of silicidation will be limited by the rate limiting step. Si and metal atoms diffuse in all phases in the system and the highest diffusion barrier controls the rate of atomic movement. These diffusion processes involve vacancies and other defects such as local amorphous disorder, grain boundaries, and interfaces which depend on the process history and are thus difficult to describe quantitatively.

As discussed earlier, the present simulations show that $Ni_2Si$ is rapidly formed in the initial stages of silicidation due to the fast diffusion of Ni in Si. There is a concomitant diffusion of Si into the metal overlayer, albeit not so rapid as the Ni diffusion in Si. The computed barrier for Si moving via a vacancy mechanism inside Ni bulk is 20 times higher than the diffusion barrier for Ni inside Si.

The simulations indicate that there is a thin layer of NiSi closest to the Si/silicide interface. In the presence of excess Si, there is a thermodynamic driving force to convert $Ni_2Si$ into NiSi. This can occur in one of two ways which both involve diffusion through the NiSi layer. Si can diffuse from the silicon substrate through NiSi and react with $Ni_2Si$ to form NiSi, via the reaction

$$Si + Ni_2Si \rightarrow 2NiSi \qquad (3)$$

The energy gain is 47 kJ/mol. The other possibility is that Ni from either the metal overlayer or from $Ni_2Si$ diffuses through NiSi to react with Si at the Si/NiSi interface. For Ni to react with Si,

$$Ni + Si \rightarrow NiSi \qquad (4)$$

the energy gain is 115 kJ/mol if Ni is taken from Ni bulk.

Diffusion of atoms through the monosilicide does not occur via an interstitial diffusion mechanism. The NiSi structure is too dense relative to the sizes of the diffusing atoms. Hence, a vacancy assisted diffusion mechanism is needed for diffusion in NiSi. The computed diffusion barriers for Ni and Pt are about 80 kJ/mol higher than the barrier for Si. Thus, the dominant diffusing species in NiSi is silicon, not nickel or platinum. A key mechanism is hence the transport of Si atoms through NiSi.

As discussed earlier, Ni preferentially occupies interstitial sites in the Si lattice and is equally stable in the two types of interstitial sites. In contrast, Pt, As, and P substitute Si atoms in the lattice. Interstitial sites are highly unfavorable for Pt, As, and P. The difference in site preference of Ni and Pt has an effect on the relative diffusion properties. The effective diffusivity of Pt in Si is much lower than that of Ni. The difference in site preference could also be related to Pt forming $Pt_2Si$ and $Pt_3Si$ structures different from the structures of Ni silicides. $Ni_3Si$ has a cubic *Pm-3m* crystal structure, while $Pt_3Si$ has an orthorhombic *Pnma* structure. $Ni_2Si$ has an orthorhombic

*Pnma* structure, while Pt$_2$Si has a tetragonal *I4/mmm* structure. This means that insertion of Pt into Ni$_3$Si and Ni$_2$Si is energetically unfavorable, since Pt prefers a different crystal structure. Pt is known to stabilize the monosilicide, which is related to the fact that NiSi and PtSi both have the same orthorhombic *Pnma* structures. PtSi has a larger volume than NiSi with about 8 % larger lattice parameters.

**D. Segregation of Pt**

Pt is found to stabilize NiSi and it is therefore desirable to add Pt in the deposition of Ni on Si. Pt is also observed to act as a diffusion barrier and to reside in Ni$_2$Si grain boundaries, thereby slowing Ni$_2$Si and Ni growth kinetics [13]. To address the question of the propensity of Pt to be in grain boundaries and interfaces, in a first step computations of surface segregation energies of Pt to NiSi and Ni$_2$Si surfaces have been performed. Segregation of alloying and doping elements to surfaces is related to trends found at grain boundaries and interfaces. The most stable low-index surfaces are the NiSi(010) surface for NiSi and the Ni$_2$Si(100) surface for Ni$_2$Si (cf. Tables VII and VIII). Both the NiSi(010) and Ni$_2$Si(100) surfaces have a mixed Ni and Si termination. The computed NiSi surface energies are in good agreement with the *ab initio* results in Ref. [22]. The energies in [22] are slightly higher (up to 6 %) than the present results, which can attributed to the use of different exchange-correlation functionals (local density approximation (LDA) in [22]).

The surface segregation energies of Pt to these surfaces are computed. Two different sites at the NiSi(010) surface are considered, namely in the first and second atomic layers closest to the free surface. The segregation energy of Pt to these sites at the NiSi(010) surface is -32 kJ/mol and -74 kJ/mol, respectively. It is thus energetically favorable for Pt to move from the interior of NiSi to a free surface. This is likely to be the case also for grain boundaries. For $Ni_2Si(100)$, segregation energies of Pt to sites in the four layers closest to the free surface have also been computed. Segregation of Pt from the interior of $Ni_2Si$ to the first, second, third, and forth Ni layers at the free surface have computed energies of -52 kJ/mol, -57 kJ/mol, -42 kJ/mol, and -26 kJ/mol, respectively. Hence it is energetically favorable for Pt to segregate also to the $Ni_2Si$ surface. As for NiSi, Pt is most stable in the second Ni layer. Segregation energies are roughly similar for NiSi and $Ni_2Si$.

Thus, there is a general trend for Pt to move from the interior of the NiSi and $Ni_2Si$ silicide phases to free surfaces and thereby most likely also to other extended planar defects like interfaces and grain boundaries. The propensity for Pt to segregate to the Si/silicide interface is investigated also for the particular case of the structure resulting from the simulated silicidation (model 1, Fig.1c). The driving force for Pt segregation is determined by swapping Pt atoms in the silicide with silicon atoms at the Si/silicide interface. This is done both for a single substitution of one Pt and Si atom, and for substitution of all five Pt atoms. The optimized atomic structures are shown in Figs. 1e and f. Substitution of a particular Pt atom (Fig. 1e) results in an energy gain of 64 kJ/mol. Substitution of five Pt atoms results in a very large energy gain of 890 kJ/mol (178 kJ/mol per Pt atom). This indicates a strong site dependence in the energy gain. There is clearly a large driving force for Pt to move from the interior of the silicides to the Si/silicide interface. Pt in grain boundaries and at interfaces may act as a diffusion barrier. The computational results are in agreement with the 3D atom probe measurements by Sonehara *et al*. showing localized Pt distributions both at the silicide surface and at the silicide/Si interface [16]. They are also consistent with a computed reduction of the silicide/Si interfacial free energy by segregated Pt [15].

The thermodynamic driving force to form platinum silicide is quite large. Reaction energies of platinum silicide formation are given in Table IX. The computations show that Pt exothermically reacts with Si to form the monosilicide PtSi. While $Pt_2Si$ might form when more Pt is supplied, where Si is available PtSi will be the stable phase formed. Hence, PtSi is a thermodynamically preferred phase in this system.

## IV. Summary and Conclusions

The present *ab initio* investigation of the formation of nickel silicides provides a consistent picture of the structure and thermodynamic properties of silicidation process and it gives clear thermodynamic data on the role of Pt atoms as stabilizer of the desired NiSi phase and as diffusion barrier. Specifically, the following picture emerges from the calculations, as illustrated in Fig. 6.

If metallic Ni is deposited on a relatively thick Si substrate, *i.e.* if the system has an excess of Si, the phase diagram and the present calculations show that the thermodynamic end-point is a two-phase system consisting of $NiSi_2$ and a Si phase with a very small amount of dissolved Ni. High process temperatures will thus drive the system towards the Si and $NiSi_2$ phases, as is in fact observed [5].

The formation of the desired monosilicide requires the steering of the silicidation process into a local minimum with a 1:1 ratio of Ni and Si atoms. In other words, the process conditions need to be chosen such that the stoichiometry of the silicide phase is controlled by the exchange between the silicide phase and a Ni-rich reservoir, either metallic Ni or $Ni_2Si$ while preventing Ni to diffuse too fast into Si or Si atoms to penetrate too quickly into the silicide and metal regions.

Keeping an excess of unreacted metal on top certainly helps to prevent the formation of NiSi$_2$. Ni has a high diffusivity inside Si, but is held back inside the silicide phases due to the high thermodynamic stability in the silicide compared with Ni residing at interstitial sites inside bulk Si.

In the initial stages of the silicidation process, Ni atoms from the Si/Ni interface will diffuse into Si and begin to transform the interface region into a silicide. Based on the present simulations, this silicide will be close to Ni$_2$Si at the Ni side and possibly there is a small region of NiSi at the silicon side. The transformation does not show a sharp breakdown of the silicon lattice, but a rather continuous enrichment of Si by interstitial Ni concurrent with displacements of Si atoms without changing the basic topology of the Si lattice. This is due to the fact that the Si sublattice in Ni$_2$Si and NiSi have the same overall topology as the pure Si lattice. In contrast to Ni atoms, which are most stable in interstitial sites in a Si lattice, Pt atoms prefer to substitute Si atoms. As a consequence, the effective diffusivity of Pt in Si is much lower than that of Ni.

Given the low diffusion barrier for Ni atoms in Si, it is likely that the overall activation barrier for the formation of Ni$_2$Si is rather low, *i.e.* the process can proceed at modest temperatures. The supply of Ni is provided by the Ni atoms at the Ni/silicide interface. The simulations show that some diffusion of Si into the Ni-rich region is possible. Hence this part of the process can be considered as an interdiffusion of the two elements, with Ni(Pt) as the main diffusing species. A sputtered layer of Ni(Pt) is likely to have a high number of defects including vacancies and internal boundaries, which facilitate the interdiffusion.

As an upper limit of diffusion barriers inside the Ni layer we can use the barrier for the hopping of a Ni atom into a neighboring vacancy. Experimentally, the monovacancy contribution to the self-diffusion in nickel single crystals has been reported as $D = 0.92 \exp(-2.88\text{eV}/kT)$ cm$^2$/s in the temperature range between 813 and 1193 K [39]. This implies an effective barrier of 2.88 eV = 278 kJ/mol. Hence, the supply of Ni atoms is likely to come mainly from interface atoms with the other Ni atoms moving collectively closer to the silicon substrate as Ni atoms diffuse away from the interface toward the silicon phase. In fact, the simulations show very clearly the shrinkage in volume during the consumption of the metallic overlayer and the formation of the silicide phase.

P and As dopants in Si are thermodynamically more stable inside the silicon phase rather than within the silicides. However, both As and P atoms are very stable at substitutional sites in metallic nickel. This means that excess metal and high processing temperatures may deplete doped silicon of P and As, if these dopant atoms can reach the Ni(Pt) metal. However, as soon as a silicide is formed between the doped silicon and the pure metal overlayer, P and As atoms are snowplowed towards the silicon substrate if the process does not proceed too fast. As discussed in Section IIIB, with very rapid quenching P and As atoms can be trapped inside the silicide layers.

The boundary between silicide and the silicon substrate remains relatively sharp (a few atomic layers) due to the fact that the chemical potential of Ni in bulk Si is 45 kJ/mol (energy required to move Ni atoms from bulk Ni into interstitial sites in silicon) whereas the formation energy of NiSi from Si and bulk Ni is -115 kJ/mol. This means that there is a strong thermodynamic driving force for Ni atoms to be inside the NiSi phase rather than being dissolved in bulk silicon. This quantifies the observation of a very low solid solubility of Ni in Si as seen in the phase diagram.

Because of the high mobility of Ni in bulk Si and the attractive chemical potential of Ni inside NiSi the concentration of Ni atoms ahead of the silicide front is determined essentially by thermodynamics, even at low temperatures in the processing. With a difference in the chemical potential of 160 kJ/mol, even at 1000 °C the equilibrium concentration of Ni in Si would only be about $7 \times 10^{-7}$ %. Upon cooling of such a sample Ni atoms would continue to diffuse from the interior of Si towards the silicide phase, and thereby reducing the amount of dissolved Ni in Si even further.

Once the $Ni_2Si$ is formed, the transformation into the desired NiSi proceeds via diffusion of Si through NiSi. This diffusion has to be facilitated by defects such as vacancies. In fact, vacancy assisted diffusion of Si atoms has a barrier 80 kJ/mol smaller than that for Ni and Pt. In other words, the dominant diffusing species in NiSi is silicon, not nickel or platinum, nevertheless the barriers are relatively high. Thus, elevated temperatures are needed to transform $Ni_2Si$ into NiSi.

The computations show a relatively high vacancy-assisted diffusivity of Ni atoms inside $Ni_2Si$, but Si atoms in this structure are rather immobile. This means that $Ni_2Si$ is a barrier for the diffusion of Si. In other words, Si atoms will diffuse through NiSi but then get stuck inside $Ni_2Si$ until this phase is transformed into NiSi. One would thus conclude that there is a reactive zone at the boundary between NiSi and $Ni_2Si$, where NiSi supplies Si atoms and the $Ni_2Si$ side provides Ni atoms. In this process, the $Ni_2Si$ lattice will eventually be depleted of Ni atoms to the point when the systems becomes actually NiSi.

The present calculations show that Pt atoms have a pronounced thermodynamic driving force to accumulate at the Si/silicide interface. Furthermore, Pt atoms have a tendency to transform $NiSi_2$ into a substitutional monosilicide of the form $Pt_xNi_{1-x}Si$. In other words, Pt atoms stabilize the desired monosilicide. This leads us to the conclusion that the diffusion of Si through NiSi is one of the key steps in the formation of Ni(Pt) monosilicide. As described in Ref. [24], the process for silicidation of Pt on Si is very similar to that of Ni. Pt atoms diffuse into Si forming $Pt_2Si$, followed by the formation of PtSi by Si diffusion through the growing monosilicide. Hence, diffusion of Si through the monosilicide (PtSi) is one of the key steps also in the formation of Si/PtSi. A similar conclusion has been reached from experimental work on the Si/CoSi system [38]. The silicidation processes with Ni and Co are again similar with regards to the main diffusing species, namely Ni (Co) atoms during formation of the $Ni_2Si$ ($Co_2Si$) and Si atoms in the transformation from $Ni_2Si$ ($Co_2Si$) to the mono-silicides NiSi and CoSi.

The calculations reveal that Pt is relatively stable in the monosilicide phase, but overall has a preference for the interfaces rather than being embedded inside single phases. As such, Pt stabilizes the monosilicide and acts as diffusion barrier preventing Si to diffuse into NiSi transforming the monosilicide into $NiSi_2$. Such a reaction would be thermodynamically favorable, albeit with a rather small reaction energy of -17.2 kJ/mol.

This work provides a detailed atomistic understanding of the structural and thermodynamic aspects of the silicidation of a Si(001) surface by Ni alloyed with Pt. The role of Pt as stabilizer of the desired nickel monosilicide has been quantified in terms of the chemical potential of Pt in the relevant silicide phases. The kinetic aspects of the silicidation process are elucidated by considering the main diffusion processes and by ranking these processes according to the corresponding activation barriers. The present results thus provide a deeper understanding and quantitative data on the nickel silicide phases, which determine the outcome of silicidation processes of silicon surface. Furthermore, *ab initio* molecular dynamics simulations show the rather high diffusivity of Ni atoms in the initial stage of silicide formation and the subsequent role of diffusing Si atoms in the formation of the desired NiSi.

# TABLES

TABLE I. Computed lattice parameters for crystalline bulk phases together with deviation from experimental values.

| System (Space Group) | Z | $a$ (Å) | $b$ (Å) | $c$ (Å) | $\Delta a$ (%) | $\Delta b$ (%) | $\Delta c$ (%) |
|---|---|---|---|---|---|---|---|
| Ni (*Fm-3m*) | 4 | 3.46 | 3.46 | 3.46 | -1.7 [40] | -1.7 | -1.7 |
| $Ni_3Si$ (*Pm-3m*) | 1 | 3.46 | 3.46 | 3.46 | -1.2 [41] | -1.2 | -1.2 |
| $Ni_{31}Si_{12}$ (*P321*) | 1 | 6.59 | 6.59 | 12.15 | -1.2 [42] | -1.2 | -1.2 |
| $Ni_2Si$ (*Pnma*) | 4 | 4.91 | 3.71 | 7.01 | -1.7 [43] | -0.3 | -0.7 |
| $Ni_3Si_2$ (*Cmc2_1*) | 16 | 12.09 | 10.70 | 6.86 | -1.1 [44] | -1.0 | -0.9 |
| NiSi (*Pnma*) | 4 | 5.08 | 3.36 | 5.53 | -1.8 [45] | 1.0 | -1.4 |
| $NiSi_2$ (*Fm-3m*) | 4 | 5.39 | 5.39 | 5.39 | -0.1 [46] | -0.1 | -0.1 |
| Si (*Fd-3m*) | 8 | 5.43 | 5.43 | 5.43 | 0.2 [47] | 0.2 | 0.2 |

TABLE II. Number density of Ni and Si atoms in silicides and Ni, Si bulk phases.

| # Density | Ni | $Ni_3Si$ | $Ni_2Si$ | NiSi | $NiSi_2$ | Si |
|---|---|---|---|---|---|---|
| Ni (at/nm$^3$) | 91.4 | 69.5 | 60.9 | 41.1 | 25.6 | 0 |
| Si (at/nm$^3$) | 0 | 23.2 | 30.4 | 41.1 | 51.3 | 50.0 |
| Total (at/nm$^3$) | 91.4 | 92.7 | 91.3 | 82.2 | 76.9 | 50.0 |

TABLE III. Energies of nickel silicide formation (kJ/mol).

| Reaction | Reaction energy | Energy per atom |
|---|---|---|
| Si + 3 Ni → $Ni_3Si$ | -206.3 | -51.6 |
| Si + 2 $Ni_3Si$ → 3 $Ni_2Si$ | -138.2 | -15.4 |
| Si + 2 Ni → $Ni_2Si$ | -183.6 | -61.2 |
| Si + $Ni_2Si$ → 2 NiSi | -47.1 | -11.8 |
| Si + Ni → NiSi | -115.4 | -57.7 |
| Si + NiSi → $NiSi_2$ | -17.2 | -5.7 |
| Ni + 2 Si → $NiSi_2$ | -132.5 | -44.2 |
| Ni + $NiSi_2$ → 2 NiSi | -98.1 | -24.5 |
| Ni + NiSi → $Ni_2Si$ | -68.3 | -22.8 |
| Ni + $Ni_2Si$ → $Ni_3Si$ | -22.7 | -5.7 |
| 3 $Ni_3Si$ + $NiSi_2$ → 5 $Ni_2Si$ | -166.6 | -11.1 |
| $Ni_2Si$ + $NiSi_2$ → 3 NiSi | -29.9 | -5.0 |

TABLE IV. Energy (in kJ/mol) required to move Ni, Pt, As, and P atoms from their elemental bulk phase into bulk Si.

|  | Ni | Pt | As | P |
|---|---|---|---|---|
| Tetrahedral site | 45 | 101 | 400 | 357 |
| 6f site | 45 | 89 | 338 | 265 |
| Substituting Si | 118 | 5 | 23 | -2 |

TABLE V. Energy (in kJ/mol) required to take Pt, As, and P atoms from their respective elemental bulk phase and substitute Ni or Si atom in silicides. A negative value indicates that the substitution is exothermic.

| System | Ni | $Ni_3Si$ | $Ni_2Si$ | NiSi | $NiSi_2$ | Si |
|---|---|---|---|---|---|---|
| Pt subst Ni | +0.26 | +18.5 | +19.5 | -21.5 | +5.5 | |
| Pt subst Si | | +271 | +206 | +148 | -82.3 | +5.4 |
| As subst Ni | -81 | +120 | +289 | +358 | +409 | |
| As subst Si | | +115 | +129 | +69 | +84 | +23.1 |
| P subst Ni | -137 | +32 | +217 | +284 | +319 | |
| P subst Si | | +56 | +41 | +9 | +52 | -2.2 |

TABLE VI. Energy difference (in kJ/mol) of As and P in bulk Si vs in silicide.

| | $Ni_3Si$ | $Ni_2Si$ | NiSi | $NiSi_2$ |
|---|---|---|---|---|
| As | 92 | 106 | 46 | 61 |
| P | 58 | 43 | 11 | 55 |

TABLE VII. $Ni_2Si$ surface energies.

| Surface | Surface energy (J/m$^2$) |
|---|---|
| $Ni_2Si$(100) Ni terminated | 2.13 |
| $Ni_2Si$(100) Si terminated | 2.64 |
| $Ni_2Si$(010) | 2.27 |
| $Ni_2Si$(001) | 2.34 |

TABLE VIII. NiSi surface energies.

| Surface | Surface energy (J/m$^2$) |
| --- | --- |
| NiSi(100) Ni terminated | 2.92 |
| NiSi(100) Si terminated | 2.42 |
| NiSi(010) | 2.19 |
| NiSi(001) Ni terminated | 2.77 |
| NiSi(001) Si terminated | 2.21 |

TABLE IX. Energies of platinum silicide formation (kJ/mol).

| Reaction | Reaction Energy |
| --- | --- |
| Pt + Si → PtSi | -146 |
| Pt + PtSi → Pt$_2$Si | -78 |
| Pt + 2Si + Ni$_2$Si → PtSi + 2NiSi | -193 |
| 2Si + Pt$_2$Si + Ni$_2$Si → 2PtSi + 2NiSi | -140 |
| 2PtSi + Ni$_2$Si → Pt$_2$Si + 2NiSi | 21 |
| 2PtSi + NiSi → Pt$_2$Si + NiSi$_2$ | 51 |

# FIGURES

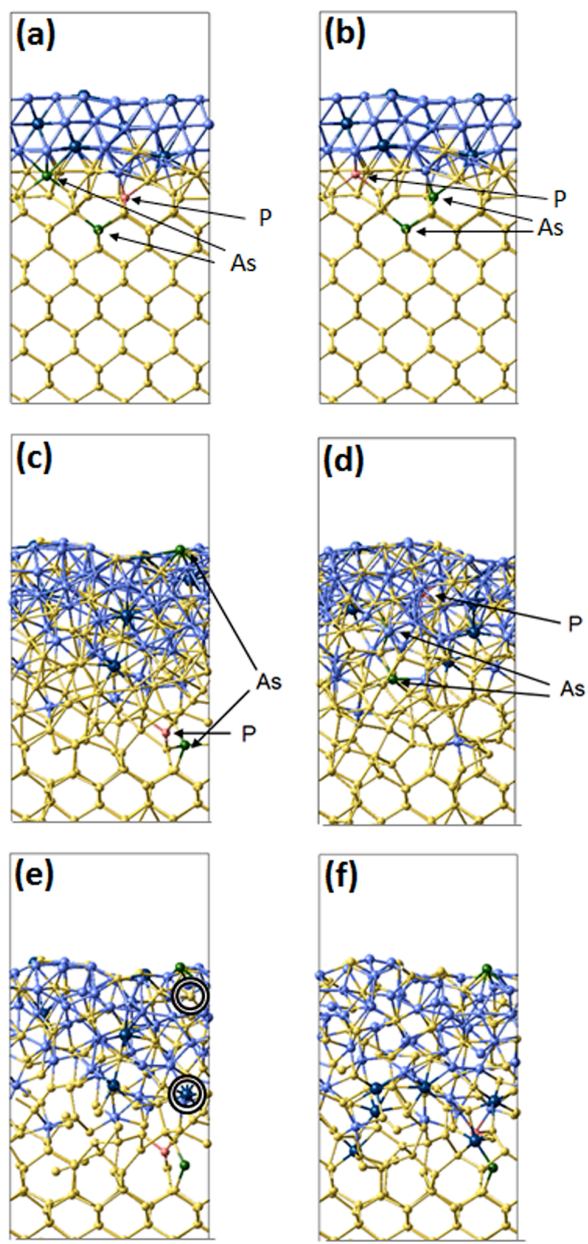

**Figure 1.** Si(001) with Ni-Pt overlayer forming a Si/silicide interface. (a) Model 1 before annealing. (b) Model 2 before annealing. (c) Model 1 after annealing. (d) Model 2 after annealing. (e) One Pt and Si atom in model 1 are swapped (encircled). (f) All Pt atoms in the silicide are swapped with Si atoms at the Si/silicide interface.

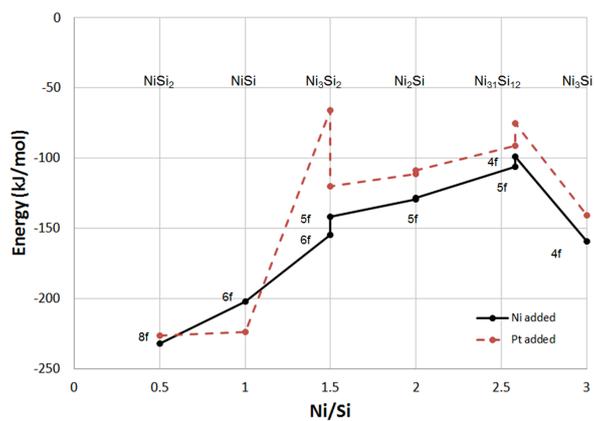

**Figure 2**. Energy gain of inserting one Ni or Pt atom into a metal-site vacancy in nickel silicide. The sites in each phase are labeled by their atomic coordination. Bulk Ni and Pt are used as reference.

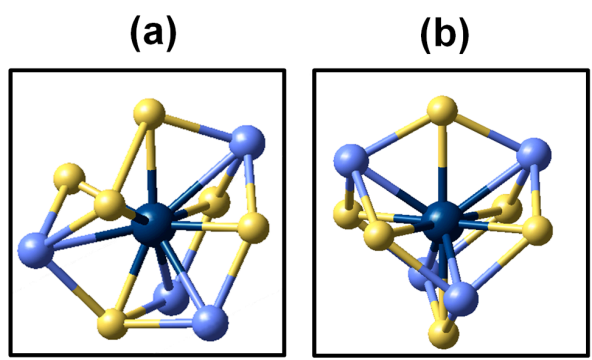

**Figure 3.** Local atomic arrangement around a Pt atom (a) substituting Ni in NiSi (b) substituting Si in $NiSi_2$.

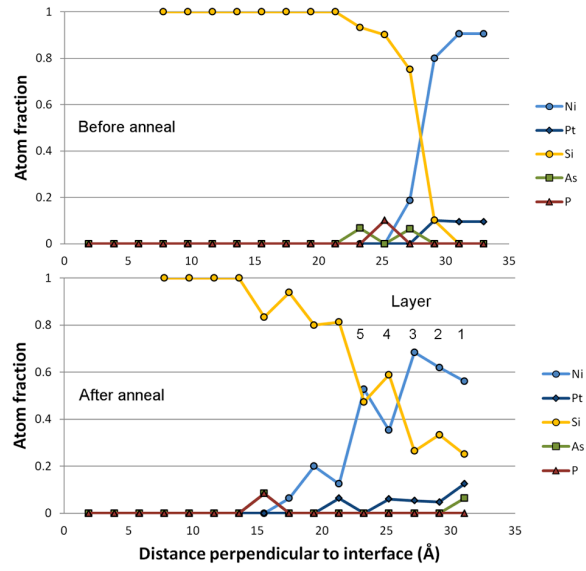

**Figure 4.** Concentration profiles of Ni, Pt, Si, As, and P before and after anneal for model 1.

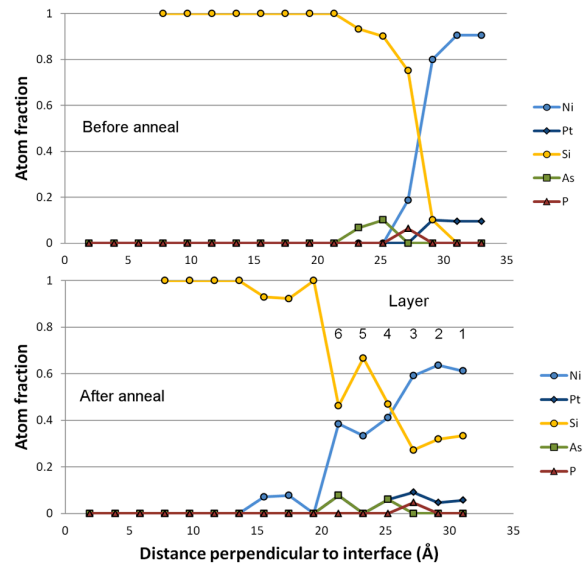

**Figure 5.** Concentration profiles of Ni, Pt, Si, As, and P before and after anneal for model 2.

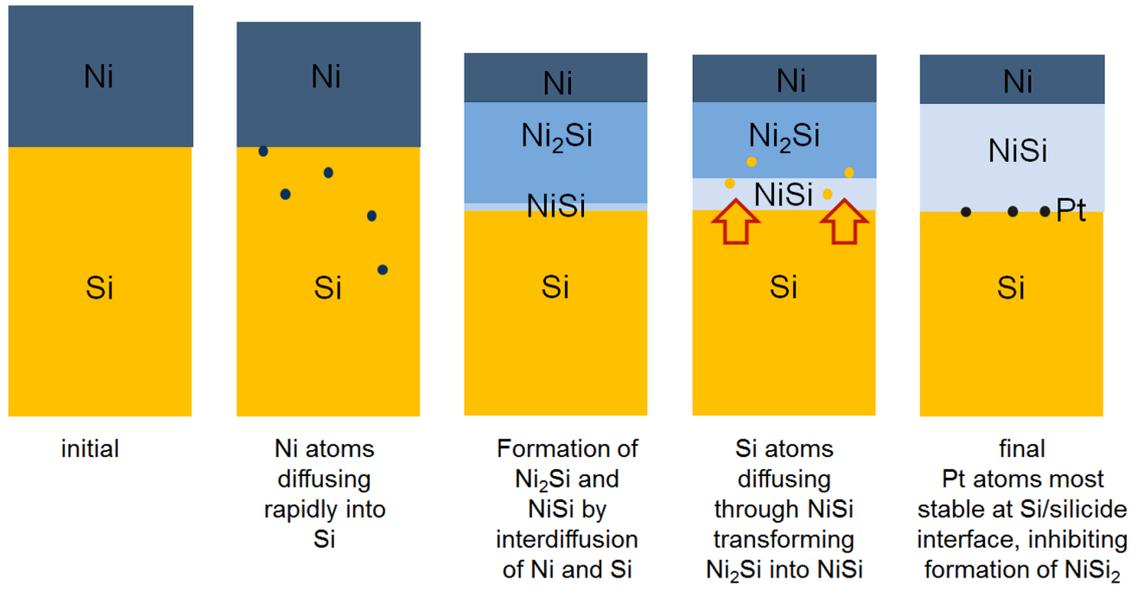

**Figure 6.** Schematics of the key steps in the formation of Ni(Pt) silicides as explained in the text.